\documentclass[12pt]{article}
\usepackage{amsmath,amssymb,amsthm,amsxtra,overpic,bbm,bm,epsfig,ulem}
\usepackage{color}
\usepackage{cite}

\usepackage{hyperref}
\usepackage{url}

\def\thefootnote{\fnsymbol{footnote}}

\begin{document}

\vspace{0.2cm}

\begin{center}
{\large\bf A direct link between unflavored leptogenesis and low-energy CP
violation via the one-loop quantum corrections}
\end{center}

\vspace{0.2cm}

\begin{center}
{\bf Zhi-zhong Xing$^{1,2}$}
and {\bf Di Zhang$^{1}$}
\footnote{E-mail: zhangdi@ihep.ac.cn (corresponding author)}
\\
{\small $^{1}$Institute of High Energy Physics and School of Physical Sciences, \\
University of Chinese Academy of Sciences, Beijing 100049, China \\
$^{2}$Center of High Energy Physics, Peking University, Beijing 100871, China}
\end{center}

\vspace{2cm}
\begin{abstract}
In the type-I seesaw mechanism the Casas-Ibarra (CI) parametrization
provides a convenient description of the Dirac neutrino mass matrix in terms of
the light and heavy Majorana neutrino masses, the lepton flavor mixing matrix
$U$ and an unknown complex orthogonal matrix $O$. If $O$ is assumed to
be real, it will be impossible to generate {\it unflavored} thermal leptogenesis
via the lepton-number-violating and CP-violating decays of the lightest heavy
Majorana neutrino. We find that this observation can be invalidated after small
but unavoidable quantum corrections to the CI parametrization are taken into
account with the help of the one-loop renormalization-group equations (RGEs)
between the seesaw and electroweak scales. We illustrate a novel and viable
unflavored leptogenesis scenario of this kind based on the RGEs in the
seesaw-extended standard model, and show its direct link to the
CP-violating phases of $U$ at low energies.
\end{abstract}

\newpage

\def\thefootnote{\arabic{footnote}}
\setcounter{footnote}{0}
\setcounter{figure}{0}

\section{Introduction}

The canonical (type-I) seesaw mechanism
\cite{Fritzsch:1975sr,Minkowski:1977sc,Yanagida:1979as,GellMann:1980vs,
Glashow:1979nm,Mohapatra:1979ia} is theoretically elegant in the sense that it
attributes the tiny masses of three known neutrinos naturally to the huge masses
of three unknown Majorana neutrinos at a sufficiently high energy scale.
Moreover, the lepton-number-violating and CP-violating decays of such
seesaw-motivated heavy neutrinos in the early Universe may provide a natural
way to account for the observed baryon-antibaryon asymmetry in
today's Universe --- a mechanism that is commonly referred to as baryogenesis
via thermal leptogenesis \cite{Fukugita:1986hr}. The key points of such a
``killing two birds with one stone" picture are briefly summarized as follows.

On the one hand, the standard model (SM) of electroweak interactions is extended
by adding three right-handed neutrino fields $N^{}_{\alpha \rm R}$
(for $\alpha = e, \mu, \tau$) and allowing lepton number violation. In this case
the gauge-invariant lepton mass terms can be written as
\begin{eqnarray}
-{\cal L}^{}_{\rm lepton} = \overline{\ell^{}_{\rm L}}
Y^{}_l H E^{}_{\rm R} + \overline{\ell^{}_{\rm L}} Y^{}_\nu
\widetilde{H} N^{}_{\rm R} + \frac{1}{2} \overline{N^{c}_{\rm R}}
M^{}_{\rm R} N^{}_{\rm R} + {\rm h.c.} \; ,
%     (1)
\end{eqnarray}
where the relevant field notations are self-explanatory, and  $M^{}_{\rm R}$ is
a symmetric Majorana mass matrix. Integrating out the heavy degrees of freedom
in Eq.~(1) \cite{Xing:2011zza}, one is left with the dimension-five
Weinberg operator ${\cal O}^{}_\nu = \left(\kappa^{}_{\alpha\beta}/2\right)
\overline{\ell^{}_{\alpha \rm L}} \tilde{H} \tilde{H}^T \ell^{c}_{\beta \rm L}$
with $\kappa = Y^{}_\nu M^{-1}_{\rm R} Y^T_\nu$ for three light neutrinos
\cite{Weinberg:1979sa}. After spontaneous gauge symmetry breaking at the
Fermi scale (i.e., $\Lambda^{}_{\rm EW} \sim 10^2$ GeV), we obtain the mass terms of
three charged leptons and three light Majorana neutrinos
\begin{eqnarray}
-{\cal L}^{\prime}_{\rm lepton} = \overline{l^{}_{\rm L}} M^{}_l E^{}_{\rm R}
+ \frac{1}{2} \overline{\nu^{}_{\rm
L}} M^{}_\nu \nu^{c}_{\rm L} + {\rm h.c.} \; ,
%     (2)
\end{eqnarray}
where $l^{}_{\rm L} = (e, \mu, \tau)^T_{\rm L}$ and
$\nu^{}_{\rm L} = (\nu^{}_e, \nu^{}_\mu, \nu^{}_\tau)^T_{\rm L}$, the
charged-lepton mass matrix $M^{}_l$ is expressed as
$M^{}_l = Y^{}_l v$ with $v \equiv \langle H^0\rangle \simeq 174$ GeV being
the vacuum expectation value of the neutral Higgs field, and the effective
Majorana neutrino mass matrix $M^{}_\nu$ is given by the famous seesaw formula
\begin{eqnarray}
M^{}_\nu = - v^2 \kappa = - M^{}_{\rm D} M^{-1}_{\rm R} M^T_{\rm D} \; ,
%     (3)
\end{eqnarray}
with $M^{}_{\rm D} = Y^{}_\nu v$ being the Dirac neutrino mass matrix. The
tiny masses $m^{}_i$ (i.e., the singular values of $M^{}_\nu$) of three light
neutrinos $\nu^{}_i$ can therefore be attributed to the large masses $M^{}_i$
(i.e., the singular values of $M^{}_{\rm R}$) of three heavy neutrinos $N^{}_i$
(for $i = 1,2,3$) as compared with $v$. In between the seesaw and Fermi scales,
which are characterized respectively by the mass $M^{}_1$ of the lightest
heavy Majorana neutrino $N^{}_1$ and the value of $v$, it is in general necessary
to consider quantum corrections to $M^{}_l$ and $M^{}_\nu$ with the help of the
renormalization-group equations (RGEs)
\cite{Chankowski:1993tx,Babu:1993qv,Antusch:2001ck,Antusch:2005gp,Mei:2005qp,
Ohlsson:2013xva}.

On the other hand, the lepton-number-violating decays
$N^{}_i \to \ell^{}_\alpha + H$ and $N^{}_i \to \overline{\ell^{}_\alpha}
+ \overline{H}$ may happen via the Yukawa interactions described by Eq.~(1).
Such processes are also CP-violating because of the interference between
their tree and one-loop amplitudes \cite{Fukugita:1986hr,Luty:1992un,Covi:1996wh,
Plumacher:1996kc}. Considering the case in which $M^{}_1 \ll M^{}_2 < M^{}_3$ holds
and all the Yukawa interactions are blind to the lepton flavors (i.e.,
the temperature of the Universe satisfies $T = M^{}_1 \gtrsim 10^{12}$ GeV \cite{Barbieri:1999ma,Endoh:2003mz,Giudice:2003jh,Nardi:2006fx,Abada:2006ea,
Blanchet:2006be}), one expects that mainly the flavor-independent (or
{\it unflavored}) CP-violating asymmetry
\begin{eqnarray}
\varepsilon^{}_{1} \equiv \frac{\displaystyle \sum^{}_\alpha \left[\Gamma
\left( N^{}_{1} \rightarrow \ell^{}_{\alpha} + H \right) - \Gamma
\left( N^{}_{1} \rightarrow \overline{\ell^{}_{\alpha}} + \overline{H} \right)\right]}
{\displaystyle \sum_\alpha \left[ \Gamma \left( N^{}_{1} \rightarrow \ell^{}_{\alpha}
+ H \right) + \Gamma \left( N^{}_{1} \rightarrow \overline{\ell^{}_{\alpha}} +
\overline{H} \right) \right]}
\simeq -\frac{3 M^{}_1}{16 \pi v^2 \left( M^{\dagger}_{\rm D}
M^{}_{\rm D}\right)^{}_{11} } \sum_i \left[ \frac{{\rm Im}
\left( M^{\dagger}_{\rm D} M^{}_{\rm D}\right)^{2}_{1i}}{M^{}_i} \right] \;
%       (4)
\end{eqnarray}
can survive and give rise to a net lepton-antilepton number asymmetry
$Y^{}_{\rm L} \equiv (n^{}_{\rm L} - n^{}_{\overline{\rm L}})/s$ with $s$ being the
entropy density of the Universe. To subsequently convert $Y^{}_{\rm L}$ to a net
baryon-antibaryon number asymmetry
$Y^{}_{\rm B} \equiv (n^{}_{\rm B} - n^{}_{\overline{\rm B}})/s$,
such an unflavored leptogenesis mechanism should keep taking effect in the
temperature range $10^2 ~{\rm GeV} \lesssim T \lesssim 10^{12} ~{\rm GeV}$
in which the non-perturbative $(B-L)$-conserving sphaleron interactions may
stay in thermal equilibrium and thus can be very efficient
\cite{Manton:1983nd,Kuzmin:1985mm,Klinkhamer:1984di}. To be explicit, we have
$Y^{}_{\rm B} = - (28/79) Y^{}_{\rm L}$ in the SM framework \cite{Kolb:1990vq,Harvey:1990qw}.
It is then possible to account for the observed baryon-to-photon ratio \cite{Aghanim:2018eyx}
\begin{eqnarray}
\eta \equiv \frac{n^{}_{\rm B}}{n^{}_\gamma} \simeq \left(6.12 \pm 0.03\right)
\times 10^{-10} \; ,
%     (5)
\end{eqnarray}
with the help of the relation $\eta = s Y^{}_{\rm B} /n^{}_\gamma \simeq 7.04 Y^{}_{\rm B}$ \cite{Xing:2011zza}. A comprehensive review of the thermal leptogenesis mechanism
with or without flavor effects can be found in
Refs.~\cite{Buchmuller:2004nz,Buchmuller:2005eh,Davidson:2008bu}.

Note, however, that the unknown flavor structure of $M^{}_{\rm D}$ is an obstacle to
the calculation of $\varepsilon^{}_1$ in Eq.~(4) \cite{Xing:2019vks}. Without invoking
any specific seesaw model and without loss of generality, one may follow Casas and
Ibarra (CI) to parametrize $M^{}_{\rm D}$
in the flavor basis where both $M^{}_l$ and $M^{}_{\rm R}$ are diagonal
(i.e., $M^{}_l = D^{}_l \equiv {\rm Diag}\{m^{}_e, m^{}_\mu, m^{}_\tau\}$ and
$M^{}_{\rm R} = D^{}_N \equiv {\rm Diag}\{M^{}_1, M^{}_{2}, M^{}_3\}$) \cite{Casas:2001sr}:
\begin{eqnarray}
M^{}_{\rm D} = {\rm i} \hspace{0.05cm} U \sqrt{D^{}_\nu} \hspace{0.1cm} O \sqrt{D^{}_N} \; ,
%     (6)
\end{eqnarray}
where $U$ is the Pontecorvo-Maki-Nakagawa-Sakata (PMNS) neutrino
mixing matrix \cite{Pontecorvo:1957cp,Maki:1962mu,Pontecorvo:1967fh}
used to diagonalize $M^{}_\nu$ in the chosen basis (i.e.,
$U^\dagger M^{}_\nu U^* = D^{}_\nu \equiv {\rm Diag}\{m^{}_1, m^{}_2, m^{}_3\}$),
and $O$ is an arbitrary complex orthogonal matrix. This CI parametrization
of $M^{}_{\rm D}$ is certainly consistent with the seesaw formula in Eq.~(3), and
it has been extensively applied to the studies of various seesaw models, leptogenesis
scenarios and lepton-flavor-violating rare decays of charged leptons.

Combining Eqs.~(4) and (6), one can immediately draw a conclusion that the unflavored
thermal leptogenesis has nothing do to with leptonic CP violation at low energies,
simply because $\varepsilon^{}_1$ is independent of $U$ (see, e.g.,
Refs.~\cite{Xing:2009vb,Rodejohann:2009cq,Antusch:2009gn} and references therein)
%%%%%%%%%%%%%%%%%%%%%%%%%%%%%%%%%%%%%%%%%%%%%%%%%%%%%%%%%%%%%%%%%%%%%%%%
\footnote{At this point one should keep in mind that the unitarity of $U$ in
the CI parametrization is consistent with the symmetry of $M^{}_\nu$ in the
leading-order seesaw formula. If a slight departure of $U$ from exact unitarity
is taken into account in the type-I seesaw mechanism, one should go beyond
Eqs.~(3) and (6) to make a self-consistent analysis of thermal leptogenesis
and its possible connection to low-energy neutrino masses, flavor mixing and
CP violation \cite{Xing:2009vb,Rodejohann:2009cq,Antusch:2009gn}.}.
%%%%%%%%%%%%%%%%%%%%%%%%%%%%%%%%%%%%%%%%%%%%%%%%%%%%%%%%%%%%%%%%%%%%%%%%
Along this line of thought, an interesting way out is to invoke flavor effects by
taking $M^{}_1 \lesssim 10^{12}$ GeV and assume $O$ to be real \cite{Pascoli:2003uh,
Pascoli:2006ie,Pascoli:2006ci,Branco:2006ce,Branco:2011zb,Moffat:2018smo}.
Then the {\it flavored} CP-violating asymmetry $\varepsilon^{}_{1 \alpha}$
(for $\alpha = e, \mu, \tau$) will depend on the CP-violating phases of $U$ in
a direct way, making it possible to connect the cosmological baryon-antibaryon
asymmetry to CP violation at low energies via flavored thermal leptogenesis.

In this paper we point out a novel way to make a direct link between {\it unflavored}
thermal leptogenesis and CP violation at low energies based on the CI parametrization
of $M^{}_{\rm D}$ and a choice of real $O$. The key point is to take into account
small but important radiative corrections to $M^{}_\nu$, which are equivalent to a
slight modification of the expression of $M^{}_{\rm D}$ on the right-hand side of
Eq.~(6), by means of the one-loop RGEs of $Y^{}_l$ and $Y^{}_\nu$ between the seesaw and
Fermi scales. In this case the low-energy PMNS matrix $U$ cannot be fully cancelled
out in the expression of $\varepsilon^{}_1$ given by Eq.~(4), and thus we are left
with an unflavored leptogenesis scenario in which the unique source of leptonic
CP violation is just the CP-violating phases of $U$.

At this point it is worth mentioning that we are motivated to assume the orthogonal
matrix $O$ to be real for two reasons. On the one hand, we intend to highlight the
novel RGE-induced effect on unflavored thermal leptogenesis, which would otherwise be
overwhelmed by those contributions originating directly from the imaginary parts of $O$.
On the other hand, switching off the imaginary parts of $O$ makes $M^{}_{\rm D}$ dependent
only upon the CP-violating phases of $U$ in the CI parametrization. This assumption
may therefore allow us to directly connect unflavored leptogenesis with low-energy
CP violation via the RGE-induced quantum corrections. But, of course, such a simple
assumption remains purely phenomenological at this stage
%%%%%%%%%%%%%%%%%%%%%%%%%%%%%%%%%%%%%%%%%%%%%%%%%%%%%%%%%%%%%%%%%%%%%%%%%%%%%%%%%%
\footnote{One may even wonder whether a viable unflavored thermal leptogenesis
scenario can be achieved in this connection by simply assuming $O$ to be the identity
matrix. We find that the answer to this question is negative, but it can be affirmative
for {\it resonant} leptogenesis with flavor effects \cite{Xing:2020ghj}.}.
%%%%%%%%%%%%%%%%%%%%%%%%%%%%%%%%%%%%%%%%%%%%%%%%%%%%%%%%%%%%%%%%%%%%%%%%%%%%%%%%%%

The remaining parts of this paper are organized as follows. In section 2 we first show
how the CI parametrization in Eq.~(6) is slightly modified by the one-loop RGE effect,
and then figure out the explicit expression of the unflavored CP violating
asymmetry $\varepsilon^{}_1$. Section 3 is devoted to illustrating that this RGE-assisted
unflavored leptogenesis scenario can work well in accounting for the observed
value of $\eta$. Finally, we make a brief summary and some concluding remarks in section 4.

\section{Quantum corrections}

\begin{figure}[h!]
  \centering
  \includegraphics[width=\linewidth]{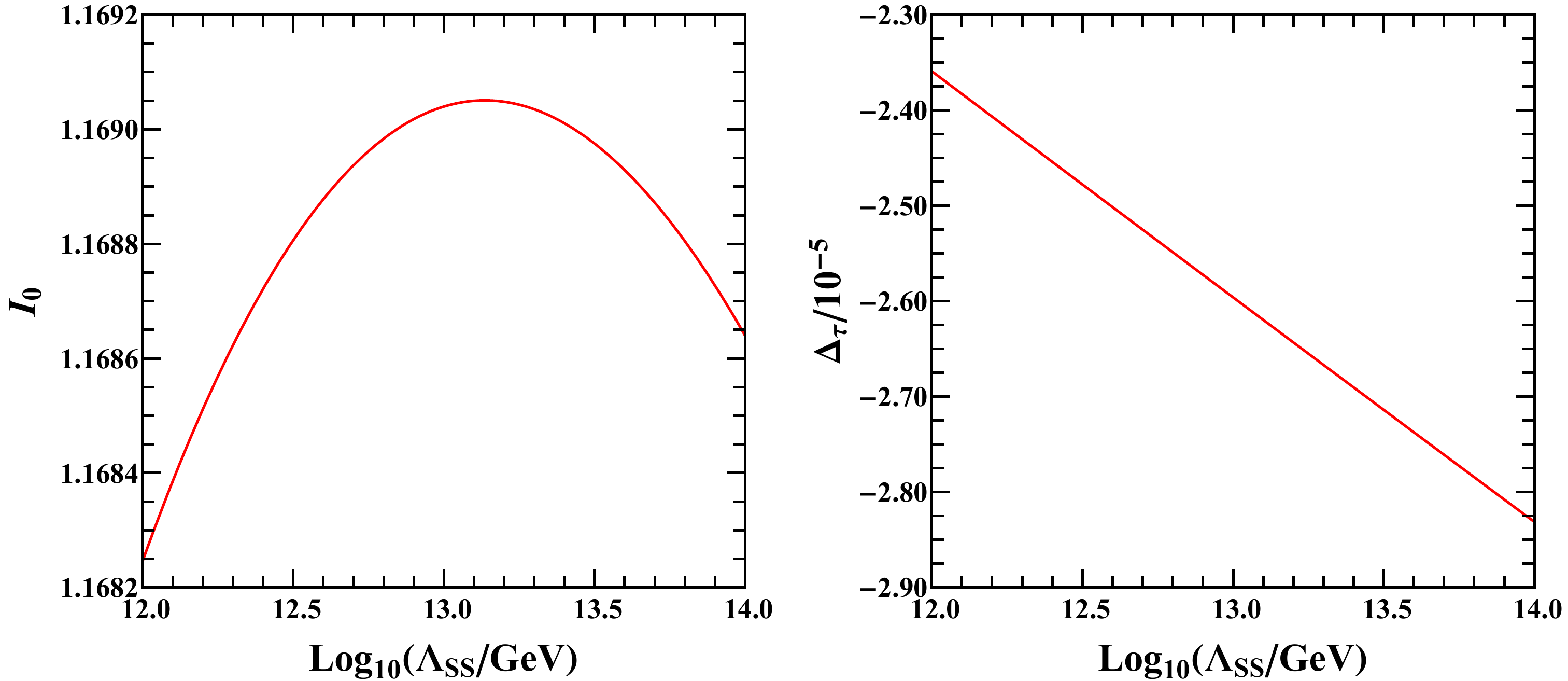}\\
  \caption{The values of $I^{}_0$ and $\Delta^{}_\tau$ against the seesaw scale
  $\Lambda^{}_{\rm SS} \in [10^{12}, 10^{14}]$ GeV in the SM.}\label{fig1}
\end{figure}

In the framework of the SM with three light Majorana neutrinos,
the one-loop RGE for the effective neutrino coupling matrix
$\kappa = Y^{}_\nu M^{-1}_{\rm R} Y^T_\nu$ evolving between the seesaw scale
($\Lambda^{}_{\rm SS} \sim M^{}_1$) and the Fermi scale ($\Lambda^{}_{\rm EW} \sim
10^2$ GeV) is given by \cite{Chankowski:1993tx,Babu:1993qv,Antusch:2001ck}
%%%%%%%%%%%%%%%%%%%%%%%%%%%%%%%%%%%%%%%%%%%%%%%%%%%%%%%%%%%%%%%%%%%%%%%%%%
\footnote{Here the coefficient of the Weinberg operator ${\cal O}^{}_\nu$ is obtained
by means of the tree-level matching condition \cite{Antusch:2005gp}.
The one-loop threshold corrections to ${\cal O}^{}_\nu$ at the matching scale 
should be taken into account in a consistent way \cite{Brivio:2018rzm},
but their effects are so small that it is absolutely safe to neglect such 
next-to-leading-order corrections in our work.}
%%%%%%%%%%%%%%%%%%%%%%%%%%%%%%%%%%%%%%%%%%%%%%%%%%%%%%%%%%%%%%%%%%%%%%%%%%%
\begin{eqnarray}
16\pi^2 \frac{{\rm d}\kappa}{{\rm d}t} = \alpha^{}_\kappa \kappa +
C^{}_\kappa \left[ \left( Y^{}_l Y^{\dagger}_l \right) \kappa +
\kappa \left( Y^{}_l Y^\dagger_l \right)^{T} \right] \;,
%   (7)
\end{eqnarray}
in which $t \equiv \ln \left(\mu/\Lambda_{\rm EW}\right)$ with $\mu$
being an arbitrary renormalization scale between $\Lambda^{}_{\rm EW}$
and $\Lambda^{}_{\rm SS}$, $C^{}_\kappa = -3/2$ and
$\alpha^{}_\kappa \approx -3g^2_2 + 6y^2_t + \lambda$ with $g^{}_2$, $y^{}_t$
and $\lambda$ standing respectively for the
${\rm SU(2)^{}_{L}}$ gauge coupling, the top-quark Yukawa coupling
and the Higgs self-coupling constant. Working in the chosen flavor basis
with both $Y^{}_l$ and $M^{}_{\rm R}$ being diagonal (i.e.,
$Y^{}_l = {\rm Diag}\{y^{}_e, y^{}_\mu, y^{}_\tau\} = D^{}_l/v$ and
$M^{}_{\rm R} = D^{}_N$), one may integrate Eq.~(7) from $\Lambda^{}_{\rm EW}$
to $\Lambda^{}_{\rm SS}$ and then arrive at
\begin{eqnarray}
\kappa \left( \Lambda^{}_{\rm SS} \right) = I^2_0 \left[
T^{}_l \cdot \kappa \left( \Lambda^{}_{\rm EW} \right) \cdot T^{}_l\right] \;,
%   (8)
\end{eqnarray}
where $T^{}_l = {\rm Diag}\{I^{}_e, I^{}_\mu, I^{}_\tau\}$, and
$I^{}_0$ and $I^{}_\alpha$ (for $\alpha = e, \mu, \tau$) are defined as
\begin{eqnarray}
I^{}_0 \hspace{-0.2cm} & = & \hspace{-0.2cm}
\exp{\left[ \frac{1}{32\pi^2} \int^{\ln{(\Lambda^{}_{\rm SS}
/\Lambda^{}_{\rm EW})}}_{0} \alpha^{}_\kappa (t) \hspace{0.05cm} {\rm d}t \right]} \;,
\nonumber \\
I^{}_\alpha \hspace{-0.2cm} & = & \hspace{-0.2cm}
\exp{\left[ \frac{C^{}_\kappa}{16\pi^2} \int^{\ln{
(\Lambda^{}_{\rm SS}/\Lambda^{}_{\rm EW})}}_{0} y^{2}_\alpha (t)
\hspace{0.05cm} {\rm d}t \right]} \; .
%   (9)
\end{eqnarray}
Given the strong hierarchy $y^2_e \ll y^2_\mu \ll y^2_\tau \ll 1$ in the
SM \cite{Xing:2019vks}, it is obvious that
$T^{}_l \simeq {\rm Diag}\{1, 1, 1+\Delta^{}_\tau\}$ holds as an
excellent approximation, where
\begin{eqnarray}
\Delta^{}_\tau = \frac{C^{}_\kappa}{16\pi^2} \int^{\ln{(\Lambda^{}_{\rm SS}
/\Lambda^{}_{\rm EW})}}_{0} y^{2}_\tau (t) \hspace{0.05cm} {\rm d}t \;
%   (10)
\end{eqnarray}
is the small $\tau$-flavored correction. The sizes of $I^{}_0$
and $\Delta_\tau$ at the seesaw scale are illustrated in Fig.~\ref{fig1}
with $\Lambda^{}_{\rm SS} \in [10^{12}, 10^{14}]$ GeV. Although the
RGE-induced effect is negligible in most cases, we are going to show that it
may play an important role in a specific unflavored thermal leptogenesis
scenario if the seesaw scale is high enough.
%%%%%%%%%%%%%%%%%%%%%%%%%%%%%%%% figure 1 %%%%%%%%%%%%%%%%%%%%%%%%%%%%%%%%%
%%%%%%%%%%%%%%%%%%%%%%%%%%%%%%%%%%%%%%%%%%%%%%%%%%%%%%%%%%%%%%%%%%%%%%%%%%%

A combination of Eqs.~(3) and (8) leads us to the RGE-corrected version of the CI
parametrization at the seesaw scale:
\begin{eqnarray}
M^{}_{\rm D} \left( \Lambda^{}_{\rm SS} \right) = {\rm i} \hspace{0.05cm}
I^{}_0 \hspace{0.05cm} T^{}_l \hspace{0.05cm} U \left( \Lambda^{}_{\rm EW} \right)
\sqrt{D^{}_\nu \left( \Lambda^{}_{\rm EW}\right)} \hspace{0.1cm} O
\sqrt{D^{}_N \left( \Lambda^{}_{\rm SS} \right)} \;,
%   (11)
\end{eqnarray}
where $U$ and $D^{}_\nu$ are the PMNS neutrino mixing matrix and the diagonal
neutrino mass matrix at low energies, respectively. Comparing Eq.~(11) with
Eq.~(6), one can see that the flavor structure of $M^{}_{\rm D}$ at $\Lambda^{}_{\rm SS}$
is slightly modified by nonzero $\Delta^{}_\tau$. As a consequence, the
product $M^\dagger_{\rm D} M^{}_{\rm D}$ becomes $U$-dependent. This new
observation motivates us to reexamine whether unflavored leptogenesis has
something to do with leptonic CP violation at low energies when $O$
is taken to be real.

To be more explicit, let us calculate the elements of
$M^\dagger_{\rm D} M^{}_{\rm D}$ that appear in Eq.~(4) at
$\Lambda^{}_{\rm SS}$ by using the RGE-corrected CI parametrization in Eq.~(11).
We obtain
\begin{eqnarray}
\left( M^\dagger_{\rm D} M^{}_{\rm D} \right)_{1i}
\hspace{-0.2cm} & = & \hspace{-0.2cm}
\left( I^2_0 \sqrt{D^{}_N} \hspace{0.1cm} O^\dagger \sqrt{D^{}_\nu} \hspace{0.1cm}
U^\dagger \hspace{0.05cm} T^2_l \hspace{0.05cm} U \sqrt{D^{}_\nu} \hspace{0.1cm} O
\sqrt{D^{}_N} \right)_{1i}
\nonumber \\
\hspace{-0.2cm} & \simeq & \hspace{-0.2cm}
I^2_0 \sqrt{M^{}_1 M^{}_i} \left[ \sum^{}_j \left(m^{}_j  O^\ast_{j1} O^{}_{ji}\right)
+ 2\Delta^{}_\tau \sum^{}_{j,k} \left(\sqrt{m^{}_j m^{}_k} \hspace{0.05cm}
O^\ast_{j1} O^{}_{ki} U^\ast_{\tau j} U^{}_{\tau k}\right) \right] +
{\cal O}(\Delta^2_\tau) \; , \hspace{0.5cm}
%   (12)
\end{eqnarray}
in which the Latin subscripts run over $(1,2,3)$, and the values of both the
neutrino masses and the PMNS matrix elements are from low energies. To minimize
the uncertainties associated with the source of CP violation, we assume the
orthogonal matrix $O$ to be real from now on. The unflavored CP-violating
asymmetry $\varepsilon^{}_1$ in Eq.~(4) turns out to be
\begin{eqnarray}
\varepsilon^{}_1 \simeq -\frac{3\Delta^{}_\tau
I^2_0 M^{}_1}{4\pi v^2} \cdot \frac{\displaystyle \sum^{}_{j>k}
\sqrt{m^{}_j m^{}_k} \left( m^{}_k - m^{}_j \right) O^{}_{j1}
O^{}_{k1} {\rm Im} \left( U^\ast_{\tau j} U^{}_{\tau k} \right)}{\displaystyle
\sum^{}_i m^{}_i O^2_{i1}} + \mathcal{O}\left(\Delta^2_\tau\right) \;.
%   (13)
\end{eqnarray}
Some immediate comments on the salient features of this result are in order.
\begin{itemize}
\item     $\varepsilon^{}_1 \propto \Delta^{}_\tau$ is naturally expected,
as one can see from Eq.~(11) in the $T^{}_l \simeq {\rm Diag}\{1, 1, 1+\Delta^{}_\tau\}$
approximation. Namely, the third row of $M^{}_{\rm D}$ is slightly corrected
due to the existence of $\Delta^{}_\tau$, so are the Yukawa coupling elements
$\left(Y^{}_\nu\right)_{\tau i}$ (for $i=1,2,3$) at the seesaw scale.
As a result, each vertex involving the $\tau$-flavored lepton doublet in the
Feynman diagrams of $N^{}_1 \to \ell^{}_\alpha + H$ and
$N^{}_1 \to \overline{\ell^{}_\alpha} + \overline{H}$ decays is slightly
modified, making it possible to trigger the interference between their
tree and one-loop amplitudes at the leading order of $\Delta^{}_\tau$
and result in the unflavored CP-violating asymmetry $\varepsilon^{}_1$ as shown
in Eq.~(13). That is why $\varepsilon^{}_1$ will automatically vanish if the RGE-induced
effect between $\Lambda^{}_{\rm EW}$ and $\Lambda^{}_{\rm SS}$ is switched off.
In other words, the unflavored CP-violating asymmetry $\varepsilon^{}_1$
is actually dependent on the $\tau$-flavored quantum correction.

\item     Eq.~(13) provides us with a direct link between unflavored leptogenesis
at the seesaw scale and the CP-violating phases of $U$ at low energies,
since $O$ has been assumed to be real. In fact, only the elements in the
first column of $O$ and those in the third row of $U$ are involved in
the expression of $\varepsilon^{}_1$. So one may simply use two rotation angles
$\theta$ and $\phi$ to make the parametrization
$(O^{}_{11}, O^{}_{21}, O^{}_{31}) = (\sin\theta\cos\phi,
\sin\theta\sin\phi, \cos\theta)$ with $\theta \in (0, \pi]$ and $\phi
\in (0, 2\pi]$. On the other hand, only two of the three CP-violating phases
of $U$ (or two combinations of theirs) take effect in $\varepsilon^{}_1$ given
by Eq.~(13)
%%%%%%%%%%%%%%%%%%%%%%%%%%%%%%%%%%%%%%%%%%%%%%%%%%%%%%%%%%%%%%%%%%%%%%%%%%
\footnote{This point will be more transparent if one adopts a particular
Euler-like parametrization of $U$ proposed by Fritzsch and one of us
\cite{Fritzsch:1997fw}, in which the
elements in the third row only involves two Majorana-type CP-violating phases.}.
%%%%%%%%%%%%%%%%%%%%%%%%%%%%%%%%%%%%%%%%%%%%%%%%%%%%%%%%%%%%%%%%%%%%%%%%%%

\item    Given the phase convention for the PMNS matrix $U$~\cite{Tanabashi:2018oca},
\begin{eqnarray}
U = \left(\begin{matrix}
c^{}_{12} c^{}_{13} & s^{}_{12} c^{}_{13} &
s^{}_{13} e^{-{\rm i} \delta} \cr
-s^{}_{12} c^{}_{23} - c^{}_{12}
s^{}_{13} s^{}_{23} e^{{\rm i} \delta} & c^{}_{12} c^{}_{23} -
s^{}_{12} s^{}_{13} s^{}_{23} e^{{\rm i} \delta} & c^{}_{13}
s^{}_{23} \cr
s^{}_{12} s^{}_{23} - c^{}_{12} s^{}_{13} c^{}_{23}
e^{{\rm i} \delta} &- c^{}_{12} s^{}_{23} - s^{}_{12} s^{}_{13}
c^{}_{23} e^{{\rm i} \delta} &  c^{}_{13} c^{}_{23} \cr
\end{matrix} \right) \left(\begin{matrix} e^{{\rm i} \rho} & 0 & 0 \cr
0 & e^{{\rm i}\sigma} & 0 \cr 0 & 0 & 1 \cr \end{matrix}\right) \; ,
%     (14)
\end{eqnarray}
in which $c^{}_{ij} \equiv \cos\theta^{}_{ij}$ and $s^{}_{ij} \equiv \sin\theta^{}_{ij}$
(for $ij = 12, 13, 23$) with $\theta^{}_{ij}$ lying in the first quadrant,
Eq.~(13) tells us that $\varepsilon^{}_1$ only contains three terms (e.g., $jk=31,32,21$)
whose Majorana CP phases are $\rho$, $\sigma$ and $\rho - \sigma$ respectively.
It is easy to see that if the transformations $\rho \to \rho \pm \pi$ and
$\sigma \to \sigma \pm \pi$ are made, either separately or simultaneously,
$\varepsilon_1$ may keep unchanged if a proper transformation of $\theta$ or $\phi$
is accordingly made (e.g., a combination of the transformations
$\rho \to \rho \pm \pi$, $\sigma \to \sigma \pm \pi$ and
$\theta \to \pi - \theta$ keeps $\varepsilon_1$ invariant). Such properties
are pretty useful for us to understand the numerical results
for the parameter space of our scenario in section 3.

\item     One may wonder what will happen if the unknown orthogonal matrix $O$
is taken to be the identity matrix. In this special case, we are left with
a much simpler result
\begin{eqnarray}
\varepsilon^{}_1 \simeq -\frac{3\Delta^{2}_\tau
I^2_0 M^{}_1}{4\pi v^2} \sum_{i} m^{}_i {\rm Im}\left(U^\ast_{\tau 1}
U^{}_{\tau i}\right)^2  + \mathcal{O} \left( \Delta^3_\tau \right) \; ,
%   (15)
\end{eqnarray}
which is proportional to $\Delta^2_\tau$ and thus strongly suppressed in
magnitude.
\end{itemize}
Needless to say, to make the unflavored leptogenesis scenario under consideration
viable in interpreting the observed baryon-antibaryon asymmetry of the Universe,
the value of $M^{}_1$ must be big enough such that both the
magnitudes of $\Delta^{}_\tau$ and $\varepsilon^{}_1$ can be properly enhanced.

\section{Unflavored leptogenesis}

Given $M^{}_1 \ll M^{}_2 < M^{}_3$ and $T = M^{}_1 \gtrsim 10^{12}$ GeV,
the Yukawa interactions described by Eq.~(1) are blind to all the lepton
flavors and thus mainly the unflavored CP-violating asymmetry $\varepsilon^{}_1$
given in Eq.~(4) survives and contributes to a net baryon-antibaryon asymmetry
via thermal leptogenesis. To be explicit, the final baryon-to-photon ratio $\eta$
is related to $\varepsilon^{}_1$ as follows \cite{Buchmuller:2002rq,Buchmuller:2003gz}:
\begin{eqnarray}
\eta \simeq -9.6 \times 10^{-3} \varepsilon^{}_1 \kappa^{}_{\rm f} \;,
%   (16)
\end{eqnarray}
where $\kappa^{}_{\rm f}$ is the efficiency factor determined by solving
of the Boltzmann equations of heavy Majorana neutrino and lepton number
densities, and it measures the washout effects caused by the
inverse decays and lepton-number-violating scattering processes. To figure out
the value of $\kappa^{}_{\rm f}$, let us first of all define the
out-of-equilibrium parameter of $N^{}_1$ decays as
$K^{}_1 \equiv \Gamma^{}_1/H(M^{}_1) = \widetilde{m}_1/m^{}_\ast $,
where $\Gamma^{}_1 = \left( Y^\dagger_\nu Y^{}_\nu \right)^{}_{11} M^{}_1 /(8\pi)$
denotes the total decay width of $N^{}_1$,
$H(M^{}_1) = \sqrt{8\pi^3 g^{}_\ast/90} M^2_1/M^{}_{\rm pl}$
is the Hubble expansion parameter at temperature $T = M^{}_1$ with $g^{}_\ast = 106.75$
being the total number of relativistic degrees of freedom in the SM and
$M^{}_{\rm pl} = 1.22 \times 10^{19} ~{\rm GeV}$ being the
Planck mass, $\widetilde{m}^{}_1 = \left( M^\dagger_{\rm D} M^{}_{\rm D}
\right)^{}_{11} / M^{}_1$ represents the effective neutrino mass, and
$m^{}_\ast = 8\pi v^2 H(M^{}_1)/M^2_1 \simeq 1.08 \times 10^{-3}$ eV stands
for the equilibrium neutrino mass. With the help of Eq.~(12), we obtain
\begin{eqnarray}
\widetilde{m}^{}_1 \simeq I^2_0 \left[ \sum^{}_i m^{}_i O^2_{i1} + 2\Delta^{}_\tau
\sum^{}_{i,j} \sqrt{m^{}_i m^{}_j} O^{}_{i1} O^{}_{j1} {\rm Re}
\left( U^\ast_{\tau i} U^{}_{\tau j} \right) \right] \; .
%   (17)
\end{eqnarray}
It is obvious that $K^{}_1$ controls whether or not the decays of
$N^{}_1$ are in equilibrium. In the far out-of-equilibrium (or weak washout) regime
(i.e., $K^{}_1 \ll 1$), $\kappa^{}_{\rm f}$ depends heavily on the initial
abundance of heavy Majorana neutrinos, and the produced $(B-L)$-asymmetry
is not reduced by washout effects. In the strong washout regime
(i.e.,$K^{}_1 \gg 1$), however, $\kappa^{}_{\rm f}$ is almost independent of
the initial conditions and hence the $(B-L)$-asymmetry produced at high temperatures
is efficiently washed out. Given the initial thermal abundance of heavy Majorana
neutrinos, the approximate analytical relation between $\kappa^{}_{\rm f}$ and
$K^{}_1$ can be expressed as \cite{Buchmuller:2004nz,Blanchet:2006be}
\begin{eqnarray}
\kappa^{}_{\rm f} \simeq \frac{2}{K^{}_1 z^{}_{\rm B} (K^{}_1)} \left[
1- \exp\left( -\frac{1}{2} K^{}_1 z^{}_{\rm B} (K^{}_1) \right) \right] \;,
%   (18)
\end{eqnarray}
with $z^{}_{\rm B} (K^{}_1) \simeq 2 + 4 K^{0.13}_1 \exp \left( -2.5/K^{}_1\right)$.
A combination of Eq.~(13) and Eqs.~(16)---(18) will therefore allow us to
estimate the value of $\eta$ via unflavored leptogenesis and examine its
dependence on the CP-violating phases of $U$ at low energies.

\begin{figure}[h!]
  \centering
  \includegraphics[width=\linewidth]{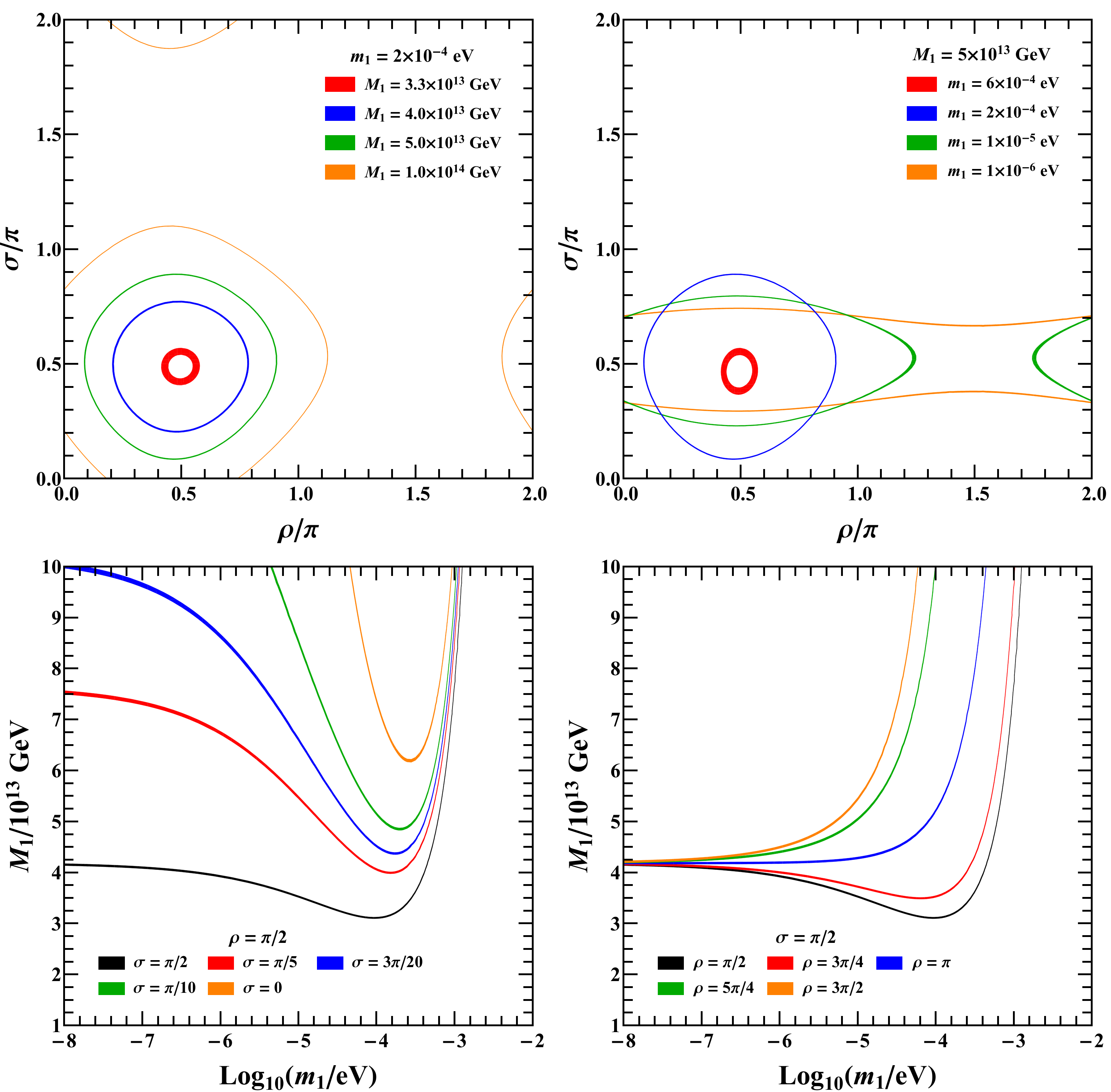}\\
  \caption{A viable unflavored leptogenesis scenario in the {\it normal} neutrino mass
  ordering case: the parameter space of $\rho$ and $\sigma$ (upper panels) with some
  given values of $m^{}_1$ and $M^{}_1$; and the parameter space of $m^{}_1$
  and $M^{}_1$ (lower panels) with some given values of $\rho$ and $\sigma$.}\label{fig2}
\end{figure}

\begin{figure}[h!]
  \centering
  \includegraphics[width=\linewidth]{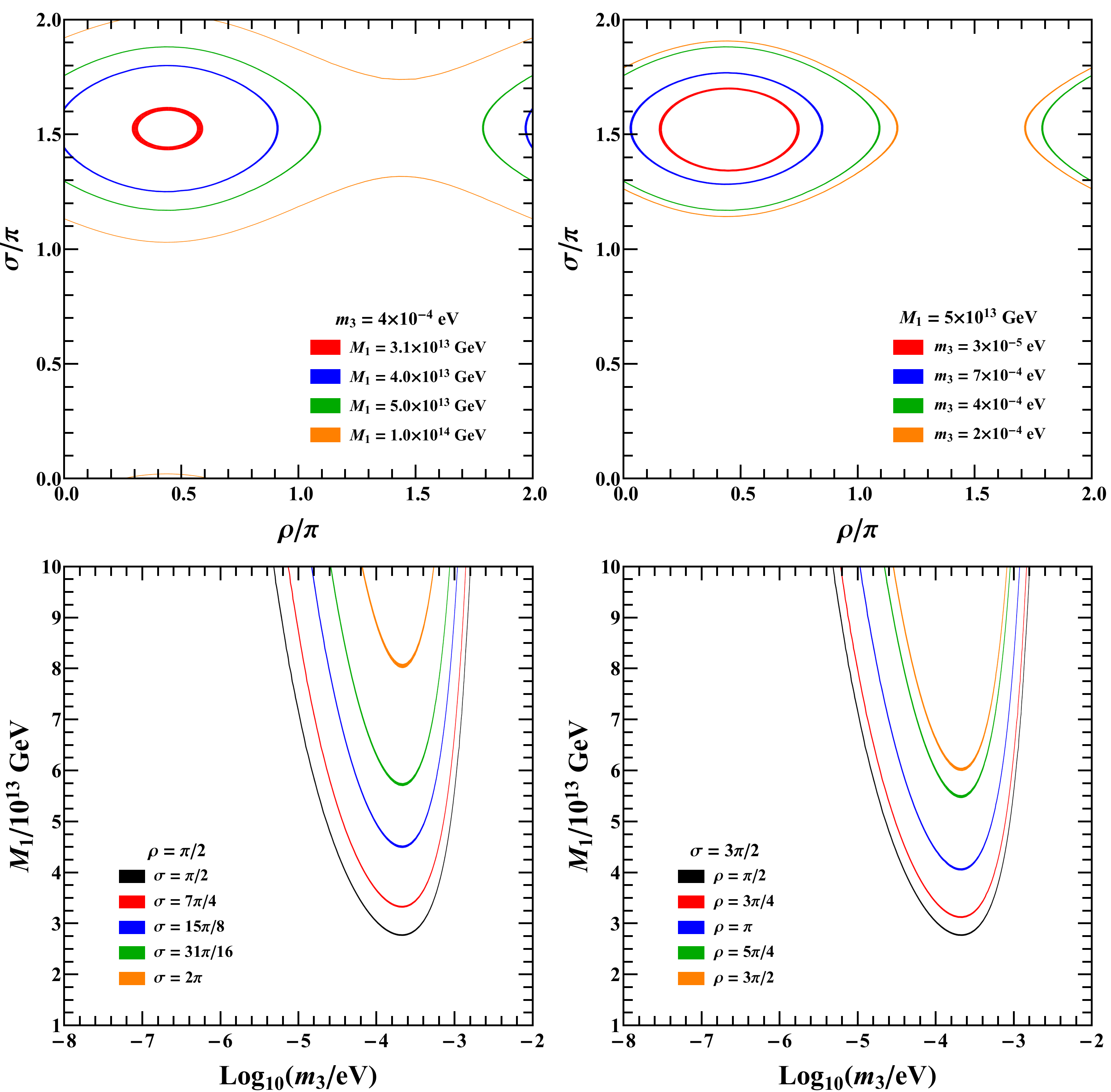}\\
  \caption{A viable unflavored leptogenesis scenario in the {\it inverted} neutrino mass
  ordering case: the parameter space of $\rho$ and $\sigma$ (upper panels) with some
  given values of $m^{}_1$ and $M^{}_1$; and the parameter space of $m^{}_1$
  and $M^{}_1$ (lower panels) with some given values of $\rho$ and $\sigma$.}\label{fig3}
\end{figure}

Adopting the parametrization of the PMNS matrix $U$ given in Eq.~(14),
we find that there are totally twelve parameters involved in our unflavored
leptogenesis scenario: the heavy neutrino mass $M^{}_1$ which determines the values
of $I^{}_0$ and $\Delta^{}_\tau$; three light neutrino masses $m^{}_i$ (for $i=1,2,3$);
three lepton flavor mixing angles $\theta^{}_{12}$, $\theta^{}_{13}$ and
$\theta^{}_{23}$; three CP-violating phases $\delta$, $\rho$ and $\sigma$;
and two free parameters $\theta$ and $\phi$ used to parametrize $O^{}_{i1}$
(for $i=1,2,3$). For simplicity, here we only input the best-fit
values of $\theta^{}_{12}$, $\theta^{}_{13}$, $\theta^{}_{23}$, $\delta$,
$\Delta m^2_{21} \equiv m^2_2 - m^2_1$ and $\Delta m^2_{31} \equiv m^2_3 - m^2_1$
(or $\Delta m^2_{32} \equiv m^2_3 - m^2_2$) extracted from a recent global analysis
of current neutrino oscillation data ~\cite{Capozzi:2018ubv,Esteban:2018azc}:
\begin{eqnarray}
\sin^2 \theta^{}_{12} = \left\{ \begin{array}{l} 0.310 \\ 0.310 \end{array} \right.
\;, \quad \sin^2 \theta^{}_{13} = \left\{ \begin{array}{l} 0.02241 \\ 0.02261
\end{array} \right. \;, \quad \sin^2 \theta^{}_{23} = \left\{ \begin{array}{l} 0.558
\\ 0.563 \end{array} \right. \;,\quad \delta = \left\{ \begin{array}{l} 222^\circ
\\ 285^\circ \end{array} \right. \;,
%   (19)
\end{eqnarray}
and
\begin{eqnarray}
\Delta m^2_{21} = \left\{ \begin{array}{l} 7.39\times 10^{-5} ~{\rm eV^2} \\
7.39\times 10^{-5} ~{\rm eV^2} \end{array} \right. \;, \quad \left\{
\begin{array}{l} \Delta m^2_{31} = 2.523 \times 10^{-3} ~{\rm eV^2} \\
\Delta m^2_{32} =  -2.509 \times 10^{-3} ~{\rm eV^2} \end{array} \right. \;,
%   (20)
\end{eqnarray}
where both the normal neutrino mass ordering (NMO, upper values) and the
inverted mass ordering (IMO, lower values) are taken into account. We also
choose a set of typical values of $\theta$ and $\phi$ in our numerical calculations:
$(\theta, \phi) = (84.9^\circ, 351.1^\circ)$ for the NMO case and
$(\theta, \phi) = (174.9^\circ, 58.0^\circ)$ for the IMO case, which both
allow $K^{}_1$ to satisfy $K^{}_1 \gtrsim 1$ if the best-fit values in Eqs.~(19)
and (20) and the reasonable ranges of $m^{}_1$ (or $m^{}_3$), $M^{}_1$, $\rho$ and $\sigma$
are input. In the $K^{}_1 \gtrsim 1$ regime any lepton-antilepton asymmetries
generated by the lepton-number-violating and CP-violating decays of $N^{}_2$
and $N^{}_3$ can be efficiently washed out, and thus we are only left with
the asymmetry produced by the decays of $N^{}_1$. The latter depends only on
four unknown parameters: $m^{}_1$ (or $m^{}_3$), $M^{}_1$, $\rho$ and $\sigma$.
In the following we shall use the observed value of $\eta$ given in Eq.~(5)
to constrain the parameter space of $\rho$ and $\sigma$ by allowing $m^{}_1$ (or $m^{}_3$)
and $M^{}_1$ to vary in the ranges $[10^{-8}, 10^{-2}]$ eV
and $[10^{13}, 10^{14}]$ GeV, respectively; or to constrain the parameter
space of $m^{}_1$ (or $m^{}_3$) and $M^{}_1$ by allowing both $\rho$ and $\sigma$
to vary in the $(0, 2\pi]$ range. Our numerical results are plotted in
Fig.~\ref{fig2} and Fig.~\ref{fig3} for the NMO and IMO cases, respectively.
Some discussions are in order.
%%%%%%%%%%%%%%%%%%%%%%%%%%%%%%%% figure 2 %%%%%%%%%%%%%%%%%%%%%%%%%%%%%%%%%
%%%%%%%%%%%%%%%%%%%%%%%%%%%%%%%%%%%%%%%%%%%%%%%%%%%%%%%%%%%%%%%%%%%%%%%%%%%
%%%%%%%%%%%%%%%%%%%%%%%%%%%%%%%% figure 3 %%%%%%%%%%%%%%%%%%%%%%%%%%%%%%%%%
%%%%%%%%%%%%%%%%%%%%%%%%%%%%%%%%%%%%%%%%%%%%%%%%%%%%%%%%%%%%%%%%%%%%%%%%%%%
\begin{itemize}
\item      {\it The NMO case}. As shown in the upper panels of Fig.~\ref{fig2}, the
parameter space of $\rho$ and $\sigma$ is mainly located in the $[0, \pi]$ range.
With $m^{}_1$ increasing or $M^{}_1$ decreasing, the values of $\rho$ and $\sigma$
will approach $\pi/2$, indicating an upper boundary for $m^{}_1$ and
a lower boundary for $M^{}_1$. Such boundaries are dominated by
$\rho = \sigma = \pi/2$, which are described by the black bands in
the lower panels of Fig.~\ref{fig2}. It is obvious that $m^{}_1 \lesssim 10^{-3}$ eV
and $M^{}_1 \gtrsim 3\times 10^{13}$ GeV hold, implying that a nearly degenerate
neutrino mass spectrum is not compatible with this unflavored thermal leptogenesis
scenario. In fact, a similar conclusion has been drawn   in Refs.~\cite{Buchmuller:2002rq,Buchmuller:2000as,Fujii:2002jw}.
We find that $m^{}_1$ may approach zero if $M^{}_1 \gtrsim 4.2\times 10^{13}$ GeV
holds. From the lower right panel of Fig.~\ref{fig2}, one can see that all the
bands converge at $M^{}_1 \simeq 4.2 \times 10^{13}$ GeV when $m^{}_1$
becomes smaller and smaller. This observation means that in the given
parameter setting the value of $\eta$ is essentially insensitive to the
value of $\rho$, as clearly shown by the yellow band with $m^{}_1=1\times10^{-6}$ eV
in the upper right panel of Fig.~\ref{fig2}. The reason is simply that
the term ${\rm Im} \left( U^\ast_{\tau 3} U^{}_{\tau2} \right)$
dominates the size of $\varepsilon^{}_1$ in Eq.~(13) because of the smallness of
$m^{}_1$. There is a similar behavior in the lower left panel of
Fig.~\ref{fig2}, in which all the bands nearly converge at
$m^{}_1 \simeq 1\times 10^{-3}$ eV when $M^{}_1 > 8\times10^{13}$ GeV holds,
and it implies that the sensitivity of $\eta$ to $\sigma$ is quite weak in this
parameter setting.

\item      {\it The IMO case}. The upper panels of Fig.~\ref{fig3} show
that the parameter space of $\rho$ and $\sigma$ is mainly located
in the $\rho\in [0,\pi]$ and $\sigma\in [\pi,2\pi]$ ranges. As in
the NMO case, $\rho$ and $\sigma$ approach $\pi/2$ and $3\pi/2$, respectively,
when $M^{}_1$ decreases. But as $m^{}_3$ increases from a small value,
$(\rho, \sigma)$ first go far away from $(\pi/2, 3\pi/2)$ and then
approach $(\pi/2, 3\pi/2)$ again, a behavior which is different from the NMO case.
Such features mean that there exit a lower boundary for $M^{}_1$ and
both lower and upper boundaries for $m^{}_3$ which are determined by $\rho=\pi/2$
and $\sigma=3\pi/2$, as explicitly shown by the black bands in the
lower panels of Fig.~\ref{fig3}. The smallest value of $M^{}_1$ needed
to fit the observed value of $\eta$ is slightly smaller than that in the NMO case,
and it is around $2.8\times10^{13}$ GeV as one can also see from the lower panels of
Fig.~\ref{fig3}. We find that $m^{}_3$ is constrained to lie in the range
$5\times10^{-6} ~{\rm eV} \lesssim m^{}_3 \lesssim 1.6\times10^{-3}$ eV when
$M^{}_1 \leq 10^{14}$ GeV is required. Both the cases of $m^{}_3 \simeq 0$ and
$m^{}_1 \approx m^{}_2 \approx m^{}_3$ are incompatible with this unflavored
thermal leptogenesis scenario for the given parameter setting.
\end{itemize}

Before ending this section, let us briefly mention an alternative possibility
that the heavy Majorana neutrinos have the initial zero abundance.
In this case the efficiency factor $\kappa^{}_{\rm f}$ is different from that
given by Eq.~(14), especially in the weak washout regime (i.e., $K^{}_1 \ll 1$)
\cite{Blanchet:2006be,Buchmuller:2004nz}. But our preliminary numerical analysis
shows that if the same values
of $\theta$ and $\phi$ are chosen to assure $K^{}_1 \gtrsim 1$, then the size
of $\kappa^{}_{\rm f}$ will be only slightly smaller than that in the situation of
the initial thermal abundance. As a consequence, the allowed lower bound of $M^{}_1$
becomes larger (i.e., $M^{}_1 \gtrsim 10^{14}$ GeV) and the behaviors of the
other three parameters (i.e., $m^{}_1$, $\rho$, $\sigma$) are quite similar to
those in the initial thermal abundance case.

\section{Summary}

With the help of the RGE-corrected CI parametrization of the Dirac neutrino mass
matrix $M^{}_{\rm D}$ in the canonical seesaw mechanism, we have shown that it is
possible to link {\it unflavored} thermal leptogenesis to CP violation at low
energies in a direct way. The point is that the PMNS matrix $U$ can no longer be
fully cancelled out in the unflavored CP-violating asymmetry $\varepsilon^{}_1$,
and thus it is the unique source of CP violation if the unknown orthogonal matrix
$O$ in the expression of $M^{}_{\rm D}$ is assumed to be real. As a numerical
exercise, we have taken special values for the elements $O^{}_{i1}$ (for $i=1,2,3$)
and adopted the best-fit values of six neutrino oscillation parameters to illustrate
the dependence of the baryon-to-photon ratio $\eta$ on both the neutrino masses (i.e.,
$M^{}_1$ and $m^{}_1$ or $m^{}_3$) and the Majorana-type CP-violating phases of $U$
(i.e., $\rho$ and $\sigma$). It is found that such a RGE-assisted unflavored
leptogenesis scenario can work well for $M^{}_1 \gtrsim 10^{13}$ GeV in the
framework of the seesaw-extended SM.

It is worth remarking that the huge gap between the seesaw scale $\Lambda^{}_{\rm SS}$
and the Fermi scale $\Lambda^{}_{\rm EW}$ makes it definitely meaningful to take into account
radiative corrections to the original CI parametrization by using the one-loop RGEs,
although such quantum effects are very small and even negligible in most cases. If one
turns to the framework of the minimal supersymmetric standard model (MSSM) extended
with the type-I seesaw mechanism, however, the RGE-induced corrections to $M^{}_{\rm D}$
are expected to be much more appreciable. In this MSSM case one may similarly explore
a direct connection between unflavored leptogenesis and CP violation at low energies
based on the RGE-corrected CI parametrization with $O$ being real \cite{Zhao2020}.

Of course, we have only focused on the effect of unflavored leptogenesis induced by
the lepton-number-violating and CP-violating decays of the lightest heavy Majorana
neutrino with the condition $10^{13} ~{\rm GeV} \lesssim M^{}_1 \ll M^{}_2 < M^{}_3$.
As far as {\it flavored} thermal leptogenesis is concerned, we find that the similar
RGE-induced effect (proportional to $\Delta^{}_\tau$) will in general become a
next-to-leading-order term in the expressions of $\varepsilon^{}_{1 \alpha}$ (for
$\alpha = e, \mu, \tau$) and hence unimportant. But such a preliminary
observation deserves a further and comprehensive study.

In short, a successful leptogenesis mechanism at the seesaw scale is generally unnecessary
to have a direct link to lepton flavor mixing and CP violation at low energies
\cite{Xing:2009vb,Buchmuller:1996pa,Davidson:2007va}, but it is always interesting to
find a specific and transparent scenario to bridge the gap between such high-scale
and low-scale physics. The present work has therefore given a new example of this kind.

\section*{Acknowledgements}

The original idea of this work came into being in a white night for one of us
(Z.Z.X.) during his participation in the workshop entitled ``New Physics on the
Low-energy Precision Frontier" at CERN, from 28 January to 7 February 2020. We
are greatly indebted to Zhen-hua Zhao for many useful discussions and a friendly
``competition", and to Shun Zhou for his enlightening comments on leptogenesis.
Our current research activities are supported in part by the National Natural
Science Foundation of China under grant No. 11775231 and grant No. 11835013.

%\vspace{0.5cm}


\newpage

\begin{thebibliography}{99}
\bibitem{Fritzsch:1975sr}
  H.~Fritzsch, M.~Gell-Mann and P.~Minkowski,
  %``Vector - Like Weak Currents and New Elementary Fermions,''
  Phys.\ Lett.\  {\bf 59B} (1975) 256.

\bibitem{Minkowski:1977sc}
  P.~Minkowski,
  %``$\mu \to e\gamma$ at a Rate of One Out of $10^{9}$ Muon Decays?,''
  Phys.\ Lett.\  {\bf 67B} (1977) 421.

\bibitem{Yanagida:1979as}
  T.~Yanagida,
  %``Horizontal gauge symmetry and masses of neutrinos,''
  Conf.\ Proc.\ C {\bf 7902131} (1979) 95.

\bibitem{GellMann:1980vs}
  M.~Gell-Mann, P.~Ramond and R.~Slansky,
  %``Complex Spinors and Unified Theories,''
  Conf.\ Proc.\ C {\bf 790927} (1979) 315
  [arXiv:1306.4669 [hep-th]].

\bibitem{Glashow:1979nm}
  S.~L.~Glashow,
  %``The Future of Elementary Particle Physics,''
  NATO Sci.\ Ser.\ B {\bf 61} (1980) 687.

\bibitem{Mohapatra:1979ia}
  R.~N.~Mohapatra and G.~Senjanovic,
  %``Neutrino Mass and Spontaneous Parity Nonconservation,''
  Phys.\ Rev.\ Lett.\  {\bf 44} (1980) 912.

\bibitem{Fukugita:1986hr}
  M.~Fukugita and T.~Yanagida,
  %``Baryogenesis Without Grand Unification,''
  Phys.\ Lett.\ B {\bf 174} (1986) 45.

\bibitem{Xing:2011zza}
  Z.~z.~Xing and S.~Zhou,
  %``Neutrinos in particle physics, astronomy and cosmology,''
  Springer-Verlag, Berlin Heidelberg (2011).

\bibitem{Weinberg:1979sa}
  S.~Weinberg,
  %``Baryon and Lepton Nonconserving Processes,''
  Phys.\ Rev.\ Lett.\  {\bf 43} (1979) 1566.

\bibitem{Chankowski:1993tx}
  P.~H.~Chankowski and Z.~Pluciennik,
  %``Renormalization group equations for seesaw neutrino masses,''
  Phys.\ Lett.\ B {\bf 316} (1993) 312
  [hep-ph/9306333].

\bibitem{Babu:1993qv}
  K.~S.~Babu, C.~N.~Leung and J.~T.~Pantaleone,
  %``Renormalization of the neutrino mass operator,''
  Phys.\ Lett.\ B {\bf 319} (1993) 191
  [hep-ph/9309223].

%\cite{Antusch:2001ck}
\bibitem{Antusch:2001ck}
  S.~Antusch, M.~Drees, J.~Kersten, M.~Lindner and M.~Ratz,
  %``Neutrino mass operator renormalization revisited,''
  Phys.\ Lett.\ B {\bf 519} (2001) 238
  [hep-ph/0108005].

\bibitem{Antusch:2005gp}
  S.~Antusch, J.~Kersten, M.~Lindner, M.~Ratz and M.~A.~Schmidt,
  %``Running neutrino mass parameters in see-saw scenarios,''
  JHEP {\bf 0503} (2005) 024
  [hep-ph/0501272].

\bibitem{Mei:2005qp}
  J.~w.~Mei,
  %``Running neutrino masses, leptonic mixing angles and CP-violating phases: From M(Z) to Lambda(GUT),''
  Phys.\ Rev.\ D {\bf 71} (2005) 073012
  [hep-ph/0502015].

\bibitem{Ohlsson:2013xva}
  T.~Ohlsson and S.~Zhou,
  %``Renormalization group running of neutrino parameters,''
  Nature Commun.\  {\bf 5} (2014) 5153
  [arXiv:1311.3846 [hep-ph]].

\bibitem{Luty:1992un}
  M.~A.~Luty,
  %``Baryogenesis via leptogenesis,''
  Phys.\ Rev.\ D {\bf 45} (1992) 455.

\bibitem{Covi:1996wh}
  L.~Covi, E.~Roulet and F.~Vissani,
  %``CP violating decays in leptogenesis scenarios,''
  Phys.\ Lett.\ B {\bf 384} (1996) 169
  [hep-ph/9605319].

\bibitem{Plumacher:1996kc}
  M.~Plumacher,
  %``Baryogenesis and lepton number violation,''
  Z.\ Phys.\ C {\bf 74} (1997) 549
  [hep-ph/9604229].

\bibitem{Barbieri:1999ma}
  R.~Barbieri, P.~Creminelli, A.~Strumia and N.~Tetradis,
  %``Baryogenesis through leptogenesis,''
  Nucl.\ Phys.\ B {\bf 575} (2000) 61
  [hep-ph/9911315].

\bibitem{Endoh:2003mz}
  T.~Endoh, T.~Morozumi and Z.~h.~Xiong,
  %``Primordial lepton family asymmetries in seesaw model,''
  Prog.\ Theor.\ Phys.\  {\bf 111} (2004) 123
  [hep-ph/0308276].

\bibitem{Giudice:2003jh}
  G.~F.~Giudice, A.~Notari, M.~Raidal, A.~Riotto and A.~Strumia,
  %``Towards a complete theory of thermal leptogenesis in the SM and MSSM,''
  Nucl.\ Phys.\ B {\bf 685} (2004) 89
  [hep-ph/0310123].

 \bibitem{Nardi:2006fx}
  E.~Nardi, Y.~Nir, E.~Roulet and J.~Racker,
  %``The Importance of flavor in leptogenesis,''
  JHEP {\bf 0601} (2006) 164
  [hep-ph/0601084].

\bibitem{Abada:2006ea}
  A.~Abada, S.~Davidson, A.~Ibarra, F.-X.~Josse-Michaux, M.~Losada and A.~Riotto,
  %``Flavour Matters in Leptogenesis,''
  JHEP {\bf 0609} (2006) 010
  [hep-ph/0605281].

\bibitem{Blanchet:2006be}
  S.~Blanchet and P.~Di Bari,
  %``Flavor effects on leptogenesis predictions,''
  JCAP {\bf 0703} (2007) 018
  [hep-ph/0607330].

\bibitem{Manton:1983nd}
  N.~S.~Manton,
  %``Topology in the Weinberg-Salam Theory,''
  Phys.\ Rev.\ D {\bf 28} (1983) 2019.

\bibitem{Klinkhamer:1984di}
  F.~R.~Klinkhamer and N.~S.~Manton,
  %``A Saddle Point Solution in the Weinberg-Salam Theory,''
  Phys.\ Rev.\ D {\bf 30} (1984) 2212.

\bibitem{Kuzmin:1985mm}
  V.~A.~Kuzmin, V.~A.~Rubakov and M.~E.~Shaposhnikov,
  %``On the Anomalous Electroweak Baryon Number Nonconservation in the Early Universe,''
  Phys.\ Lett.\  {\bf 155B} (1985) 36.

\bibitem{Kolb:1990vq}
  E.~W.~Kolb and M.~S.~Turner,
  %``The Early Universe,''
  Front.\ Phys.\  {\bf 69} (1990) 1.

\bibitem{Harvey:1990qw}
  J.~A.~Harvey and M.~S.~Turner,
  %``Cosmological baryon and lepton number in the presence of electroweak fermion number violation,''
  Phys.\ Rev.\ D {\bf 42} (1990) 3344.

\bibitem{Aghanim:2018eyx}
  N.~Aghanim {\it et al.} [Planck Collaboration],
  %``Planck 2018 results. VI. Cosmological parameters,''
  arXiv:1807.06209 [astro-ph.CO].

\bibitem{Buchmuller:2004nz}
  W.~Buchmuller, P.~Di Bari and M.~Plumacher,
  %``Leptogenesis for pedestrians,''
  Annals Phys.\  {\bf 315} (2005) 305
  [hep-ph/0401240].

\bibitem{Buchmuller:2005eh}
  W.~Buchmuller, R.~D.~Peccei and T.~Yanagida,
  %``Leptogenesis as the origin of matter,''
  Ann.\ Rev.\ Nucl.\ Part.\ Sci.\  {\bf 55} (2005) 311
  [hep-ph/0502169].

\bibitem{Davidson:2008bu}
  S.~Davidson, E.~Nardi and Y.~Nir,
  %``Leptogenesis,''
  Phys.\ Rept.\  {\bf 466} (2008) 105
  [arXiv:0802.2962 [hep-ph]].

\bibitem{Xing:2019vks}
  Z.~z.~Xing,
  %``Flavor structures of charged fermions and massive neutrinos,''
  arXiv:1909.09610 [hep-ph].

\bibitem{Casas:2001sr}
  J.~A.~Casas and A.~Ibarra,
  %``Oscillating neutrinos and $\mu \to e, \gamma$,''
  Nucl.\ Phys.\ B {\bf 618} (2001) 171
  [hep-ph/0103065].

\bibitem{Pontecorvo:1957cp}
  B.~Pontecorvo,
  %``Mesonium and anti-mesonium,''
  Sov.\ Phys.\ JETP {\bf 6} (1957) 429
   [Zh.\ Eksp.\ Teor.\ Fiz.\  {\bf 33} (1957) 549].

\bibitem{Maki:1962mu}
  Z.~Maki, M.~Nakagawa and S.~Sakata,
  %``Remarks on the unified model of elementary particles,''
  Prog.\ Theor.\ Phys.\  {\bf 28} (1962) 870.

\bibitem{Pontecorvo:1967fh}
  B.~Pontecorvo,
  %``Neutrino Experiments and the Problem of Conservation of Leptonic Charge,''
  Sov.\ Phys.\ JETP {\bf 26} (1968) 984
   [Zh.\ Eksp.\ Teor.\ Fiz.\  {\bf 53} (1967) 1717].

\bibitem{Xing:2009vb}
  Z.~z.~Xing,
  %``Casas-Ibarra Parametrization and Unflavored Leptogenesis,''
  Chin.\ Phys.\ C {\bf 34} (2010) 1
  [arXiv:0902.2469 [hep-ph]].

\bibitem{Rodejohann:2009cq}
  W.~Rodejohann,
  %``Non-Unitary Lepton Mixing Matrix, Leptogenesis and Low Energy CP Violation,''
  EPL {\bf 88} (2009) no.5,  51001
  [arXiv:0903.4590 [hep-ph]].

\bibitem{Antusch:2009gn}
  S.~Antusch, S.~Blanchet, M.~Blennow and E.~Fernandez-Martinez,
  %``Non-unitary Leptonic Mixing and Leptogenesis,''
  JHEP {\bf 1001} (2010) 017
  [arXiv:0910.5957 [hep-ph]].

\bibitem{Pascoli:2003uh}
  S.~Pascoli, S.~T.~Petcov and W.~Rodejohann,
  %``On the connection of leptogenesis with low-energy CP violation and LFV charged lepton decays,''
  Phys.\ Rev.\ D {\bf 68} (2003) 093007
  [hep-ph/0302054].

\bibitem{Pascoli:2006ie}
  S.~Pascoli, S.~T.~Petcov and A.~Riotto,
  %``Connecting low energy leptonic CP-violation to leptogenesis,''
  Phys.\ Rev.\ D {\bf 75} (2007) 083511
  [hep-ph/0609125].

\bibitem{Pascoli:2006ci}
  S.~Pascoli, S.~T.~Petcov and A.~Riotto,
  %``Leptogenesis and Low Energy CP Violation in Neutrino Physics,''
  Nucl.\ Phys.\ B {\bf 774} (2007) 1
  [hep-ph/0611338].

\bibitem{Branco:2006ce}
  G.~C.~Branco, R.~Gonzalez Felipe and F.~R.~Joaquim,
  %``A New bridge between leptonic CP violation and leptogenesis,''
  Phys.\ Lett.\ B {\bf 645} (2007) 432
  [hep-ph/0609297].

\bibitem{Branco:2011zb}
  G.~C.~Branco, R.~G.~Felipe and F.~R.~Joaquim,
  %``Leptonic CP Violation,''
  Rev.\ Mod.\ Phys.\  {\bf 84} (2012) 515
  [arXiv:1111.5332 [hep-ph]].

\bibitem{Moffat:2018smo}
  K.~Moffat, S.~Pascoli, S.~T.~Petcov and J.~Turner,
  %``Leptogenesis from Low Energy $CP$ Violation,''
  JHEP {\bf 1903} (2019) 034
  [arXiv:1809.08251 [hep-ph]].

\bibitem{Xing:2020ghj}
  Z.~z.~Xing and D.~Zhang,
  %``Bridging resonant leptogenesis and low-energy CP violation with an RGE-modified seesaw relation,''
  Phys.\ Lett.\ B {\bf 804} (2020) 135397
  [arXiv:2003.06312 [hep-ph]].

\bibitem{Brivio:2018rzm}
  I.~Brivio and M.~Trott,
  %``Examining the neutrino option,''
  JHEP {\bf 1902} (2019) 107
  [arXiv:1809.03450 [hep-ph]].

\bibitem{Fritzsch:1997fw}
  H.~Fritzsch and Z.~Z.~Xing,
  %``Flavor symmetries and the description of flavor mixing,''
  Phys.\ Lett.\ B {\bf 413} (1997) 396
  [hep-ph/9707215].

\bibitem{Tanabashi:2018oca}
  M.~Tanabashi {\it et al.} [Particle Data Group],
  %``Review of Particle Physics,''
  Phys.\ Rev.\ D {\bf 98} (2018) no.3,  030001.

\bibitem{Buchmuller:2002rq}
  W.~Buchmuller, P.~Di Bari and M.~Plumacher,
  %``Cosmic microwave background, matter - antimatter asymmetry and neutrino masses,''
  Nucl.\ Phys.\ B {\bf 643} (2002) 367
   Erratum: [Nucl.\ Phys.\ B {\bf 793} (2008) 362]
  [hep-ph/0205349].

\bibitem{Buchmuller:2003gz}
  W.~Buchmuller, P.~Di Bari and M.~Plumacher,
  %``The Neutrino mass window for baryogenesis,''
  Nucl.\ Phys.\ B {\bf 665} (2003) 445
  [hep-ph/0302092].

\bibitem{Capozzi:2018ubv}
  F.~Capozzi, E.~Lisi, A.~Marrone and A.~Palazzo,
  %``Current unknowns in the three neutrino framework,''
  Prog.\ Part.\ Nucl.\ Phys.\  {\bf 102} (2018) 48
  [arXiv:1804.09678 [hep-ph]].

\bibitem{Esteban:2018azc}
  I.~Esteban, M.~C.~Gonzalez-Garcia, A.~Hernandez-Cabezudo, M.~Maltoni and T.~Schwetz,
  %``Global analysis of three-flavour neutrino oscillations: synergies and tensions in the determination of $\theta_{23}$, $\delta_{CP}$, and the mass ordering,''
  JHEP {\bf 1901} (2019) 106
  [arXiv:1811.05487 [hep-ph]].

\bibitem{Buchmuller:2000as}
  W.~Buchmuller and M.~Plumacher,
  %``Neutrino masses and the baryon asymmetry,''
  Int.\ J.\ Mod.\ Phys.\ A {\bf 15} (2000) 5047
  [hep-ph/0007176].

\bibitem{Fujii:2002jw}
  M.~Fujii, K.~Hamaguchi and T.~Yanagida,
  %``Leptogenesis with almost degenerate majorana neutrinos,''
  Phys.\ Rev.\ D {\bf 65} (2002) 115012
  [hep-ph/0202210].

\bibitem{Zhao2020}
  Z.~h.~Zhao,
  %``Renormalization group assisted leptogenesis in the minimal type-I seesaw model,"
  arXiv:2003.00654 [hep-ph].

\bibitem{Buchmuller:1996pa}
  W.~Buchmuller and M.~Plumacher,
  %``Baryon asymmetry and neutrino mixing,''
  Phys.\ Lett.\ B {\bf 389} (1996) 73
  [hep-ph/9608308].

\bibitem{Davidson:2007va}
  S.~Davidson, J.~Garayoa, F.~Palorini and N.~Rius,
  %``Insensitivity of flavoured leptogenesis to low energy CP violation,''
  Phys.\ Rev.\ Lett.\  {\bf 99} (2007) 161801
  [arXiv:0705.1503 [hep-ph]].

\end{thebibliography}
\end{document}